\documentclass[journal=jacsat,manuscript=article]{achemso}

\usepackage[version=3]{mhchem} 
\usepackage{color}
\usepackage{amsmath}
\usepackage{booktabs}
\usepackage{tablefootnote}

\newcommand*\mcii[1]{\multicolumn{2}{c}{#1}}
\newcommand*\mciii[1]{\multicolumn{3}{c}{#1}}
\newcommand*\F[1]{\mathbf{F}^\text{#1}}

\author{Chenghan Li}
\affiliation[Caltech]
{Division of Chemistry and Chemical Engineering, California Institute of Technology, Pasadena, CA, 91125, USA}
\alsoaffiliation[Marcus]
{Marcus Center for Theoretical Chemistry, California Institute of Technology, Pasadena, CA, 91125, USA}
\author{Garnet Kin-Lic Chan}
\email{gkc1000@gmail.com}
\affiliation[Caltech]
{Division of Chemistry and Chemical Engineering, California Institute of Technology, Pasadena, CA, 91125, USA}
\alsoaffiliation[Marcus]
{Marcus Center for Theoretical Chemistry, California Institute of Technology, Pasadena, CA, 91125, USA}

\title[Machine Learning Hamiltonian]
  {Predictive Free Energy Simulations Through Hierarchical Distillation of Quantum Hamiltonians}


\begin{document}

%
%
%
%
%

\begin{abstract}
Obtaining the free energies of condensed phase chemical reactions remains computationally prohibitive for high-level quantum mechanical methods. We introduce a hierarchical machine learning framework that bridges this gap by distilling knowledge from a small number of high-fidelity quantum calculations into increasingly coarse-grained, machine-learned quantum Hamiltonians. By retaining explicit electronic degrees of freedom, our approach further enables a faithful embedding of quantum and classical degrees of freedom that captures long-range electrostatics and the quantum response to a classical environment to infinite order. As validation, we compute the proton dissociation constants of weak acids and the kinetic rate of an enzymatic reaction entirely from first principles, reproducing experimental measurements within chemical accuracy or their uncertainties. Our work demonstrates a path to condensed phase simulations of reaction free energies at the highest levels of accuracy with converged statistics.
\end{abstract}



\section{Main}

Free energies are the driving force for numerous chemical and biochemical phenomena, but their accurate computation in the condensed phase presents a grand challenge for theory. Bridging the gap between the quantum mechanics of electrons and the macroscopic time and length scales on which real-world (bio-)chemical processes occur requires both high-fidelity electronic structure methods and extensive statistical thermal sampling. Although classical molecular dynamics (MD) simulations using empirical force fields (FFs) can reach long timescales, particularly when combined with specialized hardware\cite{shaw2021anton}, these potentials cannot reliably model chemical bond-breaking and forming events. Conversely, high-level quantum chemistry methods that accurately describe electron correlation are limited to small systems, with recent efforts with the gold-standard coupled cluster method and accurate basis sets reaching only picoseconds of dynamics for 19 atoms in the gas phase, even with the use of cost-reduction strategies that utilize the local nature of electron correlation\cite{zhang2024performant}.

Machine learning (ML) potentials, which regress quantum ground-state energies from molecular geometries, have emerged as a promising strategy to bridge this gap,  but they still face several critical challenges. First, most modern ML potentials rely on equivariant message-passing to achieve accuracy but are highly demanding in terms of computation and memory for large-scale problems\cite{qiao2022informing,Batatia2022mace,musaelian2023learning,batatia2025design,wood2025family,kang2025orbitall}. Second, ML potentials (especially neural-network-based variants) are often data-hungry, requiring large datasets that are prohibitively expensive to generate using high-level quantum chemistry. These two challenges are particularly pronounced in condensed-phase simulation due to the exponentially large configurational and chemical space that must be statistically sampled for computing thermal averages and to be well-represented in the training data. A multi-scale approach, similar to the hybrid quantum mechanics/molecular mechanics (QM/MM) method but replacing QM with ML, represents a natural solution but introduces a third challenge: standard ML potentials lack explicit electronic degrees of freedom, making it difficult to model the crucial response of the ML-described subsystem to the long-range electrostatics generated by the classical environment. To address this, one must either rely on a computationally inexpensive but potentially inaccurate physical model to interact with and respond to the MM potential\cite{wu2017internal,zhang2018solvation,böselt2021machine,zeng2021development,giese2025transferability,kim2021doubly,snyder2022facilitating,hofstetter2022graph,galvelis2023nnp,snyder2023bridging,kalayan2024neural,grisafi2024accelerating,zinovjev2024emle,bensberg2025machine,bensberg2025hierarchical,semelak2025advancing,sha2025modeling,novacek2025pm6} and/or modify the ML architecture to accept the MM information as additional ML input\cite{zhang2018solvation,böselt2021machine,zeng2021development,giese2025transferability,pan2021machine,hofstetter2022graph,lei2024learning,mazzeo2024electrostatic,xie2025multiscale,bensberg2025machine,bensberg2025hierarchical,song2025nepoip}


Here, we introduce a hierarchical Hamiltonian learning framework that provides a unified solution to the above challenges. Instead of directly regressing the potential energy from atomic coordinates, our approach retains explicit electronic degrees of freedom by systematically coarse-graining and parameterizing different levels of \emph{quantum} Hamiltonians. This bottom-up strategy begins by distilling the energy information from a small number of high-accuracy wavefunction quantum chemistry calculations into a cheaper, custom-parameterized, density functional theory (DFT). The learnt Kohn-Sham functional is then applied in the DFT/MM framework to generate a larger, condensed-phase dataset, which in turn is used to train a final, highly efficient machine-learned semi-empirical (SEQM) Hamiltonian\cite{dral2015machine,pan2021machine,zhou2022deep,fan2022obtaining,sun2023machine,hu2023treating,gu2024deep,soccodato2024machine,suman2025exploring,fan2025advancing} embedded within a parameterizable classical environment (ML SEQM/MM). This hierarchical structure brings two key advantages: first, it
enables data-efficient learning from the ground up, and second, 
by targeting explicit electronic representations, it provides a physically rigorous, non-perturbative framework for ML/MM coupling that naturally captures long-range electrostatics. 


From a technical perspective, our work builds on recent progress made by some of us as well as from the literature. The accurate wavefunction quantum chemistry data 
utilizes our implementation of differentiable local coupled cluster theory~\cite{nagy2019approaching,zhang2024performant} to generate energies and forces at gold-standard accuracy for systems with more than 40 atoms using accurate basis sets. Our DFT/MM simulations use our GPU implementation with multipole acceleration of electrostatics~\cite{li2025accurate} to generate condensed phase data with the proper treatment of long-range electrostatics. Finally, our learning framework utilizes both differentiable Kohn-Sham functionals implemented in this work as well as differential semi-empirical quantum models from the literature~\cite{friede2024dxtb}, combined in a new ML SEQM/MM setup that incorporates pre-trained equivariant models for feature extraction~\cite{kovacs2025mace}.

We demonstrate our approach in the context of two challenging motivating applications. The first, the proton dissociation of weak amino acids (lysine (Lys) and aspartate (Asp)), serves as a model of proton transport in the condensed phase, and features long-range charge separation and significant, non-trivial solvent reorganization. Our framework now enables us to study the free energy profile of proton dissociation with explicitly QM modeled regions with more than two hundred atoms (embedded in the classical environment). We show that this potential of mean force yields the absolute p$K_\text{a}$ of the weak acids, independent of any experimental data, to leading accuracy. Our second, the catalysis of the Claisen rearrangement by chorismate mutase (CM), is a prototypical enzyme reaction featuring non-trivial electronic rearrangement in the presence of a complex environment.
Our hierarchical setup now allows us to obtain reaction kinetics on converged potential energy surfaces with good control of the statistical thermal sampling, recovering the experimental rate constant to within chemical accuracy. Together, these demonstrate the potential of hierarchical Hamiltonian learning as a path to condensed phase simulations of free energies and kinetics based on the highest accuracy quantum chemistry data in simulations with converged statistics.

\section{Results}

\begin{figure}
    \centering
    \includegraphics[width=0.8\linewidth]{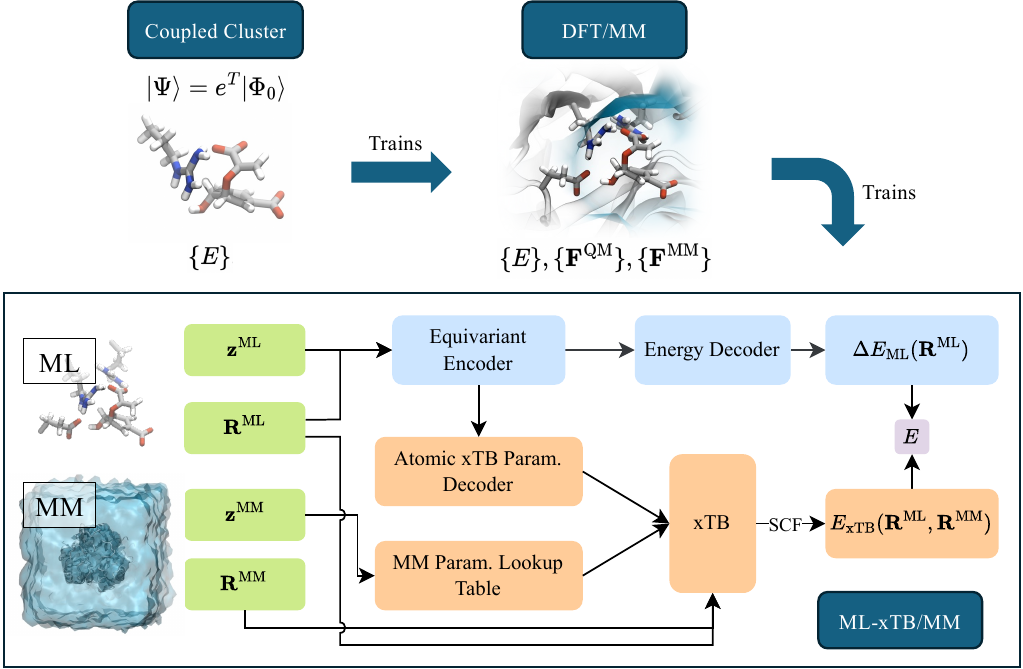}
    \caption{Model architecture. Knowledge distillation starts from high-level quantum calculations on small clusters in the gas phase, that is then distilled into a density functional theory. The DFT-based QM/MM generates new data in the condensed phase to train a machine-learned semi-empirical Hamiltonian (ML-xTB; framed). In the architecture of ML-xTB, the green blocks represent the inputs with $\mathbf{z}$ being the element types and $\mathbf{R}$ being the atomic coordinates. The blue blocks represent a neural-network-based featurizer and a neural-network potential as a dispersion correction. The orange blocks represent the tight-binding parameter predictor, MM charge, and radius lookup table, and the ground state solver.}
    \label{fig:arch}
\end{figure}

We began by calculating the energies and forces for geometries obtained from enhanced sampling simulations of the reactions of interest, using a version of local natural orbital coupled cluster singles and doubles with perturbative triples, LNO-CCSD(T)\cite{rolik2011general,zhang2024performant} (see SI Data Preparation for details of the enhanced sampling and coupled cluster calculations). We used large computational basis sets (triple-zeta and quadruple-zeta) to extrapolate to the complete basis limit and carefully characterized the convergence of the local truncation error. We find that the local correlation error mainly contributes to a global shift in the energy (see SI Error Analysis). As this does not affect the (free) energy differences, the more important error to assess is that associated with non-parallelity, or the energy differences between configurations. We estimate this energy difference error from canonical CCSD(T)/CBS to be very small ($\sim 0.2$ kcal/mol or less, as detailed in SI Error Analysis). Note that directly using such high-level quantum calculations to drive MD simulations is completely impractical for obtaining meaningful statistics, and is expensive for generating a large amount of ML training data -- even using our efficient implementation of LNO-CCSD(T)\cite{zhang2024performant} on truncated atom clusters (containing 32 atoms for Asp/Lys and 43 atoms for CM; see SI for more details) a single energy calculation for the largest QM region took approximately 1 day on 1 CPU node. Since we aimed to use only modest levels of computation, we generated only O(10)-O(100) reference data points at this level in this work.

\begin{table}[H]
    \centering
{\small
    \begin{tabular}{lccccccccc}
    \toprule
                             &\mcii{Asp/Lys cluster} &\mciii{Asp/Lys QM/MM}            &  CM cluster   &\mciii{CM QM/MM}                \\
    \midrule      
                             &  $E$ &  $\F{QM}$      &  $E$ & $\F{QM}$ & $\F{MM}$      &  $E$          &  $E$ & $\F{QM}$ & $\F{MM}$                  \\
                              \cmidrule(lr){2-3}       \cmidrule(lr){4-6}              \cmidrule(lr){7-7}      \cmidrule(lr){8-10}
    $\omega$B97X-3c          & 0.73 &        1.7     &      &          &               & 1.9           &      &          &                           \\
\textbf{reparametrized DFT}  & 0.40 &        0.95    &      &          &               & 0.28          &      &          &                           \\
    \midrule      
    Fine-tuned MACE-OFF23    &      &                & 2.4  & 0.77     &    25         &               &\mciii{OOM}                                  \\
    DPRc                     &      &                & 1.2  & 1.3      &    2.8        &               & 2.2  &  1.1     &  1.6                      \\
    \textbf{ML-xTB}          &      &                & 1.0  & 0.9      &    2.4        &               &0.95  &  0.74    &  1.6                      \\
    \midrule             
    GFN1-xTB response        &      &                & 3.5  & 2.0      &    3.8        &               &2.1   &  1.5     &  1.8                      \\
    \bottomrule
    \end{tabular}
    }
    \caption{Model accuracy in terms of mean absolute errors in energy $E$ (kcal/mol), QM atom forces $\mathbf{F}^\mathrm{QM}$ and MM atom forces $\mathbf{F}^\mathrm{MM}$ (kcal/mol/\AA). The MM force errors are normalized by the number of QM atoms instead of the number of MM atoms for more informative numbers. Models developed in this work are highlighted in bold in the first column. Errors of DFT are computed from the reference gas-phase LNO-CCSD(T) data. Errors of ML-xTB/MM are computed from the reparametrized DFT QM/MM. The GFN1-xTB response energy is defined as $\Delta E=E(\text{QM/MM})-E(\text{QM})$ and $\Delta\mathbf{F}=\mathbf{F}(\text{QM/MM})-\mathbf{F}(\text{QM})$ and the error is obtained by comparing to the reparametrized DFT $\Delta E$ and $\Delta\mathbf{F}$. For the trained models, i.e. reparametrized DFT, ML-xTB, DPRc, and MACE-OFF23, the errors are measured on the validation sets. For models not trained or reparameterized ($\omega$B97X-3c and GFN1-xTB) the errors are measured on the full data. See text for definitions of acronyms and models.}
    \label{tab:errors}
\end{table}

\begin{figure}
    \centering
    \includegraphics[width=0.85\linewidth]{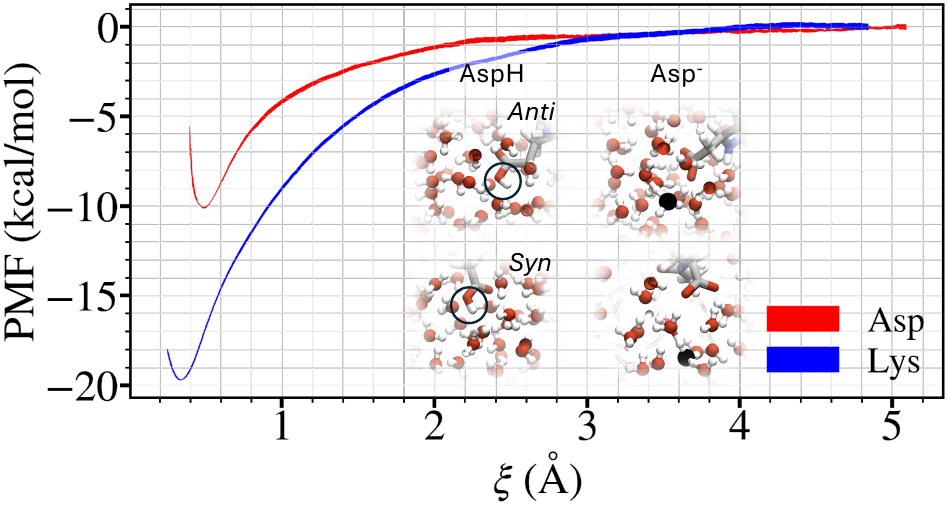}
    \caption{Potentials of mean force of proton dissociation from Asp and Lys as a function of the center of excess charge distance from their titratable moieties. Typical molecular configurations of protonated Asp (ApsH) and the deprotonated form (Asp$^-$) are shown in the insets. QM atoms are shown as opaque, while the MM atoms are shown as transparent (note only MM atoms near the QM atoms are visible). In AspH, there are \textit{syn} and \textit{anti} conformations corresponding to distinct proton positions (circled). In Asp$^-$, the hydronium, \ce{H3O+} (its oxygen shown by a black sphere) can also take \textit{syn}- and \textit{anti}-like positions. }
    \label{fig:proton}
\end{figure}

The next step in our approach (Fig.~\ref{fig:arch}) is to distill the CCSD(T) energy surface into a coarse-grained quantum Hamiltonian, for which we adopted a Kohn-Sham Hamiltonian ansatz, specifically the $\omega$B97X-3c density functional (DF) form, extended to contain 17 parameters (see SI DFT Parametrization and DFT/MM Data Generation). We then
re-parameterized the functional using the CCSD(T) energies (and no other electronic properties). This training was made feasible by a gradient-based optimization leveraging our GPU implementation of DFT\cite{wu2024python,li2025introducing}, and analytic functional derivatives with respect to the parameters, developed in this work. Choosing a DF imposes a very strong constraint on our model compared to a free-form neural network potential; thus, high data efficiency is expected in this step. Indeed, we found that 10-100 CCSD(T) energies were already sufficient to train a robust DF. For example, a DF trained on proton dissociation data from Asp achieved a comparable level of accuracy for Lys (training energy MAE 0.47 kcal/mol, and force MAE 1.1 kcal/mol/\AA, compared to the validation errors in Table~\ref{tab:errors}), and a DF trained on small atom clusters (the substrate and R90 side chain) of the CM reaction correctly predicted the energies (within 0.28 kcal/mol MAE) of much larger clusters (the substrate, R90, R7 and E78). 

Using GPU-accelerated QM/MM DFT\cite{li2025accurate}, we were then able to generate significantly more data ($\sim$ 10 times more), as well as extend the length scale from the above atom clusters to larger  
QM atom clusters (43 atoms for Asp, 46 atoms for Lys, and 72 atoms for CM) embedded in a full condensed-phase environment: MM water, as well as the protein in the case of CM, described by a standard empirical force field (see SI for details), including full periodic electrostatics. With this ability to generate highly accurate quantum data in the condensed phase, there remains the challenge of training an ML model given the mixed-resolution nature of the QM/MM data and the large total system size ($\sim$10000 atoms for Asp/Lys and $\sim$50000 atoms for CM). We found that the smallest one of a series of pre-trained foundation FF models, MACE-OFF23(S)\cite{kovacs2025mace}, could barely fit the Asp/Lys system into an Nvidia A100 GPU's memory (using $\sim$70 GB of the total 80 GB), while the larger ones could not fit at all. Even the MACE-OFF small model could not fit into memory when used for the entire CM system (denoted OOM in Table~\ref{tab:errors}). As a baseline, we fine-tuned the MACE-OFF23(S) model on the QM/MM Asp and Lys energy and forces, but this did not yield satisfactory accuracy on the MM atoms, shown by the 25 kcal/mol/\AA~of MM force MAE, and also reflected by the large energy MAE (Table~\ref{tab:errors}).

To address these challenges, we trained an even more coarse-grained semi-empirical quantum Hamiltonian from the DFT/MM energies and forces, taking a self-consistent-charge tight-binding Hamiltonian, GFN1-xTB\cite{grimme2017robust,friede2024dxtb}, as our Hamiltonian ansatz. Due to the minimal basis and the approximations for electron correlation and electrostatics in GFN1-xTB, it is generally not quantitatively accurate without re-parameterization. Crucially, the GFN1-xTB model cannot respond in the same way as DFT to the MM electrostatic potential, with an energy MAE larger than 2 kcal/mol (see GFN1-xTB Response in Table~\ref{tab:errors}). This means that even if we perfectly corrected the GFN1-xTB gas-phase energies and forces using an ML-FF within a $\Delta$-learning framework as recently proposed\cite{snyder2022facilitating,snyder2023bridging,novacek2025pm6}, it would still fail to accurately describe the condensed phase. 
This deficiency motivates our approach, in which we trained an ML-predicted GFN1-xTB Hamiltonian to correctly respond to the MM long-range electrostatics while adding an ML-potential-based dispersion correction acting solely among QM atoms. A critical component of our approach is that the ground-state potential energy surface where the MD evolves is computed from the self-consistent-field iterations of the ML-xTB Hamiltonian, and thus the response to the MM potential is captured to infinite order, in contrast to finite-order corrections based on atomic charges, polarizabilities\cite{kim2021doubly,galvelis2023nnp,kalayan2024neural,zinovjev2024emle,bensberg2025machine,semelak2025advancing,sha2025modeling}), and the QM electron density\cite{grisafi2024accelerating}. In our architecture, we employed a pre-trained equivariant graph neural network, MACE-OFF24(M)\cite{kovacs2025mace} as the featurizer and appended individual prediction heads to predict the xTB Hamiltonian parameters, a dispersion energy correction for the QM region. The MM charges and radii entering into the QM/MM electrostatic interaction were also trainable, in a geometry-independent manner (the MM charges and radii from the empirical force field were retained for the pure MM interactions, see SI ML-xTB).
Importantly, only the QM atoms are visible to the featurizer, and it (and the xTB parameter prediction head) must learn to modulate the response of the xTB Hamiltonian to the external MM potential, without direct knowledge of the MM coordinates. We found that this architecture successfully achieved chemical accuracy in terms of validation MAE (Table~\ref{tab:errors}). As another baseline, we trained a range-corrected deep potential (DPRc)\cite{zeng2021development}, which does not parameterize the xTB Hamiltonian itself, but learns a force-field correction to GFN1-xTB/MM using both the QM atoms and MM atoms as input. We found this yielded larger energy and force errors than our only-QM-visible approach (Table~\ref{tab:errors}), especially in the more complicated CM case.

\begin{table}[H]
    \centering
    \begin{tabular}{cccc}
    \toprule
         &  Theory          & \multicolumn{2}{c}{Expt.}             \\
    \midrule
    Asp  &  $3.7\pm0.1^a$   &   3.8$^b$                 & \\
    Lys  &  $10.5\pm0.1^a$  &   11.2$^b$                & \\
    \bottomrule
    \end{tabular}
    \caption{p$K_\text{a}$ from ML-xTB/MM MD and experiments.}
$^a$ statistical errors from block averaging.
$^b$ corrected for nuclear quantum effects and ionic strength from the raw values 3.67 and 10.40 measured by potentiometric titrations\cite{thurlkill2006pk}; see SI Experimental p$K_a$ Processing.\\

    \label{tab:pka}
\end{table}

\begin{figure}
    \centering
    \includegraphics[width=0.3\linewidth]{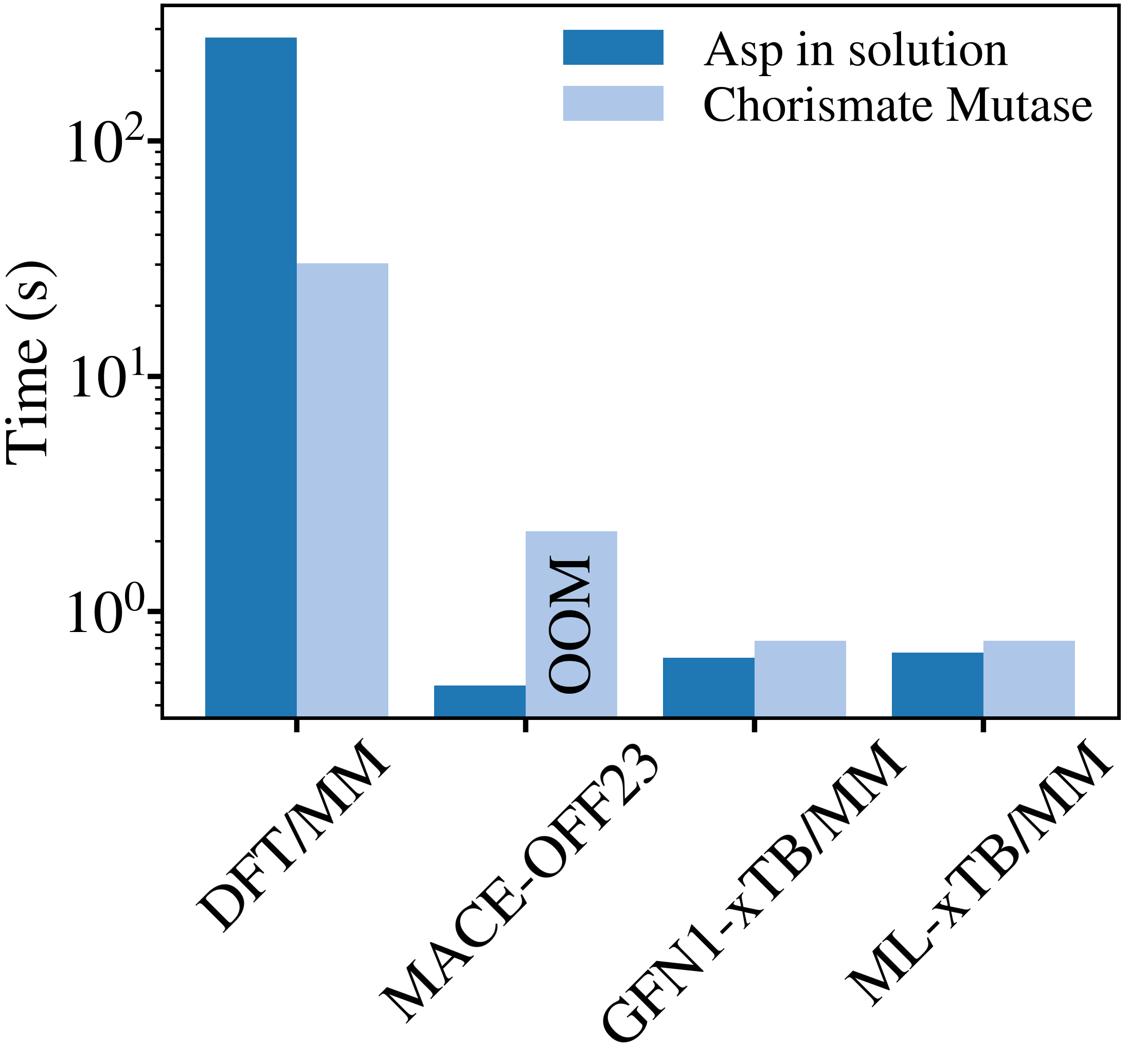}
    \caption{Wall time of one MD step on one A100 GPU of an Nvidia DGX100. The MACE-OFF23(S) model ran out of memory for the CM system, and the time was estimated assuming linear scaling in system size. The DFT, GFN1-xTB, and ML-xTB simulations correspond to a QM subsystem embedded in a MM environment, while the MACE-OFF23(S) model simulates the whole system with the ML-FF. The QM regions were one aspartate and 64 water molecules for the Asp in solution, and the substrate, the R90, R7, and E78 residue side chains for CM.}
    \label{fig:timing}
\end{figure}

We next consider the performance of these ML-xTB/MM models in full enhanced sampling MD simulations to compute key observables in our two target applications.
For the proton dissociation of Asp and Lys in water, we used the ML-xTB/MM model in conjunction with a large QM region containing the full amino acid and 64 nearby water molecules, in total more than two hundred atoms. Such a large QM region is crucial to accommodate the solvation of the excess proton in its dissociated limit, $\sim$5~\AA~away from the titratable moieties: the Asp carboxylic group and the Lys amine group. We employed replica-exchange umbrella sampling\cite{sugita2000multidimensional}, biasing the distance of the excess proton (tracked by the center of excess charge\cite{li2021using}) from the titratable groups, and computed the potential of mean force (PMF) of the dissociation reactions. To handle the statistical indistinguishability of QM and MM waters, we used the FIRES restraint (see SI FIRES~\cite{rowley2012solvation}) which is exact if the QM and MM descriptions give the same potential energy surface. 
The efficiency of our ML-xTB/MM (400-fold faster than DFT/MM; see Fig.~\ref{fig:timing}) enabled us to run nanosecond-long trajectories where each replica visited every umbrella window at least once. This is needed to sample both the \textit{syn} and \textit{anti} conformations of a protonated Asp (Fig.~\ref{fig:proton} inset left column), which are separated by a high free energy barrier in the protonated state\cite{lim2019assessing}. The \textit{syn}$\leftrightarrow$\textit{anti} transition is assisted by the exchanges between protonated and deprotonated umbrella windows, the latter of which more easily samples different proton positions relative to the carboxylic oxygen through proton Grotthuss hopping\cite{de1806decomposition} among waters (Fig.~\ref{fig:proton} inset right column). The resulting PMFs show one single well corresponding to the protonated Asp/Lys states, highlighting their weak acid nature. To validate our results, we computed the p$K_\text{a}$ values of the two residues by integrating the PMF over the protonated well (see SI for more details) and found excellent agreement with experimental measurements to within chemical accuracy (1 kcal/mol = 0.73 pH unit at 298.15K). For context, we note that the best performing absolute p$K_\text{a}$ prediction methods require experimental input for related compounds and/or the     proton solvation free energy~\cite{bochevarov2016multiconformation,yu2018weighted,luo2024bridging,bergazin2021evaluation}, and in predictions on drug-like small molecules achieve at best a similar accuracy to our result~\cite{bergazin2021evaluation}. Perhaps the most comparable theoretical approaches are those that use a free energy perturbation cycle that computes the acid deprotonation free energy\cite{riccardi2005p,li2003pka,li2025general} and obtain the absolute p$K_\text{a}$ through an independent estimate of the proton solvation free energy~\cite{malloum2021determination}, which carries an uncertainty $>$~1 kcal/mol. 

\begin{table}[H]
    \centering
    \begin{tabular}{ccccc}
    \toprule
                             &  \multicolumn{3}{c}{Theory}               &    Expt.    \\
    \midrule
                             &  DFT                       & \multicolumn{2}{c}{ML-xTB} &             \\
    \cmidrule(lr){2-2}\cmidrule{3-4}
 $V_f^\text{max}$ (kJ/mol)   &  65                        &      65     & 59           &             \\
 $k_\text{cat} $ (s$^{-1}$)  & 1.1$\pm$0.2                & 1.9$\pm$0.5 & 1.5$\pm$0.8  &  16$\pm$14  \\
    \bottomrule
    \end{tabular}
    \caption{Catalytic rate constant of chorismate mutase from \textit{Bacillus subtilis}. Theoretical errors are statistical errors among 11 flooding runs. The DFT results were extracted from previous work\cite{li2025accurate}. The raw experimental value was taken from Ref.~\cite{kast1996chorismate}, and corrected for nuclear quantum effects and temperature\cite{li2025accurate}.}
    \label{tab:kcat}
\end{table}

For our prototype enzymatic reaction, the CM-catalyzed chorismate-to-prephenate transformation, we ran 
conformational flooding simulations\cite{grubmuller1995predicting} to 
compute the  rate constant $k_\text{cat}$. Unlike the proton dissociation reaction, this reaction is a local chemical transformation, but involves a more complex electronic process (a concerted pericyclic reaction) and takes place in a heterogeneous and complicated environment. Due to the non-trivial electronic structure, standard density functionals (such as a range-separated hybrid DFT) cannot describe the energetics to within chemical accuracy (Table~\ref{tab:errors}). The training of the functional in our framework is thus necessary even for a qualitative description of the kinetics. In earlier work, we studied this reaction using GPU accelerated DFT/MM flooding MD and a  $\omega$B97X-3c functional reparametrized to accurate reaction barrier energetics~\cite{li2025accurate}. These simulations (for a specific binding mode) achieved  good agreement with experiment in the rate constant, but required a large flooding level $V_f^\text{max}=65$ kJ/mol to enhance the barrier crossing sufficiently for practical DFT/MM simulations. Indeed, such 
a high flooding level is only $\sim$1 kcal/mol lower than the forward reaction free energy barrier\cite{li2025accurate}, and thus a chemical reaction is typically observed within 20 picoseconds of simulation, but, at this level of flooding, one cannot guarantee that the core assumption of flooding based simulations, namely unperturbed dynamics around the transition state, is satisfied. As a first test, we ran flooding MD with ML-xTB/MM using the same maximum flooding level $V_f^\text{max}=65$ kJ/mol as our previous DFT/MM with a similarly revised $\omega$B97X-3c functional, accumulating statistics over 11 independent flooding runs and a total of $\sim$200 picoseconds accumulated sampling. The resulting $k_\text{cat}$ agrees very well with the DFT/MM result (Table~\ref{tab:kcat}), while with ML-xTB/MM we achieved the same amount of sampling with a 40-fold speed-up (Fig.~\ref{fig:timing}).
Importantly, however, the greater efficiency of the ML-xTB/MM model enabled us to run an order of magnitude longer trajectories ($\sim$2 nanoseconds) with a lower flooding level ($V_f^\text{max}=59$ kJ/mol). This meant that we could converge the sampling of the reactant basin--a critical requirement to obtain a reliable rate constant. Using this order of magnitude increase in sampling, we found $k_\text{cat}$ to be well converged with respect to $V_f^\text{max}$ (Table~\ref{tab:kcat}). We can thus conclude that the agreement between the theoretical and experimental rate constant is not accidental (the main remaining unquantified uncertainty is from the size of the QM region). Indeed, using both the converged potential energy surface and converged sampling, the deviation from experiment is within the range of chemical accuracy 
(1 kcal/mol = 5.4 fold in rates at 300 K) as would be expected from an accurate model.

\section{Discussion}

In summary, we have proposed a hierarchical machine learning strategy that is initiated with a small amount ($O(10-100)$) of high accuracy data and then propagates this information across space and time scales to successfully simulate complex condensed-phase chemical reactions with free energies converged to $\sim$1~kcal/mol and/or rate constants approaching the experimental uncertainty.
Crucially, rather than training ML potentials directly, our approach is based on training a hierarchy of ML quantum Hamiltonians, with a final embedding in empirical force fields. We showed that this approach benefits from physical constraints and an explicit treatment of electronic structure provided by the quantum Hamiltonian (as opposed to free-form ML potentials that lack explicit electrons), to provide a unified solution to challenges associated with data scarcity, long-range electrostatics, and the computational efficiency of learning and inference in large-scale condensed phase problems. 
Although we only used modest computational resources for simulation and data generation in this work, an interesting future direction is to utilize this same hierarchical framework in conjunction with active learning, for example, to augment both the high-level quantum chemistry wavefunction data and the reparametrized DFT energies and forces on ML-xTB sampled geometries. This would provide an efficient way to test and ensure convergence with respect to 
training data size. We anticipate this approach will be particularly valuable for even more challenging problems, such as catalysis in metalloenzymes, where the complicated electronic structure invalidates standard parameterized DFT\cite{zhai2023multireference}, and may require expensive high-level quantum reference calculations beyond  coupled-cluster theory. The developments in our work suggest that, even in such complicated chemical reactions, statistical sampling of free energies and kinetics at ambient temperature may soon be conceivable.

\begin{acknowledgement}
This work was primarily supported by the US Department of Energy, Office of Science, Basic Energy Sciences, through  Award No. DE-SC0023318. GKC acknowledges additional support in the conceptualization phase from the Dreyfus Foundation, under the program Machine Learning in the Chemical Sciences and Engineering, and from the Simons Investigator program. The authors thank Hezhou Zhang for fitting the center of excess charge parameters for the lysine molecule.

%

\end{acknowledgement}


\bibliography{main}

\end{document}


\tableofcontents



\section{Abbreviations}
We summarize the abbreviations used.
\begin{itemize}
    \item 1-RDM -- one-body reduced density matrix
    \item Asp -- aspartic acid
    \item ASPC -- always stable predictor-corrector
    \item BFGS -- Broyden–Fletcher–Goldfarb–Shanno
    \item CBS -- complete basis set
    \item CCSD(T) -- coupled cluster with singles, doubles, and perturbative triples
    \item CEC -- center of excess charge
    \item CM -- chorismate mutase
    \item DFT -- density functional theory
    \item FIRES -- flexible inner region ensemble separator
    \item HF -- Hartree-Fock
    \item LNO -- local natural orbital
    \item Lys -- lysine
    \item MAE -- mean absolute error
    \item MD -- molecular dynamics
    \item ML -- machine learning
    \item MP2 -- second-order Møller–Plesset perturbation theory
    \item NEB -- nudged elastic band
    \item OPES -- on-the-fly probability enhanced sampling
    \item PDB -- protein data bank
    \item PES -- potential energy surface
    \item PMF -- potential of mean force
    \item QM/MM -- hybrid quantum mechanics / molecular mechanics
    \item REUS -- replica-exchange umbrella sampling
    \item SCF -- self consistent field
    \item SMD -- steered molecular dynamics
    \item WHAM -- weighted histogram analysis method
    \item xTB -- extended tight binding
\end{itemize}

\section{Data Preparation}

\subsection{MAE Definition}

We use the following MAE definition for energy differences throughout the main text and this document, unless noted otherwise. For the energy, the MAE is defined as
\begin{align}
    \frac{1}{\sum_M N_M}\sum_{M} \sum_{i\in M} |E_i-E_i^\text{ref} - \frac{1}{N_M} \sum_{j\in M} (E_j-E_j^\text{ref})|
    \label{eq:mae}
\end{align}
where $\text{ref}$ denotes the reference value, $M\in\{\text{Asp},\text{Lys},\text{CM}\}$, $N_M$ is the number of Asp/Lys/CM geometries in the data set, and $i$ and $j$ index a data point (geometry). We thus only measure the energy differences between geometries, but not between different chemical compositions; this convention is consistent for all the models and methods we evaluated. For forces, MAE is defined as
\begin{align}
    \frac{1}{3 N_\text{atom}} \sum_{a}\sum_{u\in{x,y,z}} |F_{au}-F_{au}^\text{ref}|
\end{align}
where $a$ indexes an atom and $N_\text{atom}$ is the total number of atoms in the data set, summing over all geometries.

\subsection{Initial QM/MM Sampling}
\subsubsection{Asp and Lys}
The initial structures of solvated Asp and Lys were prepared using the CHARMM-GUI\cite{jo2008charmm}. One protonated amino acid in the form of \ce{Ac-X-NH2} (X$=$Asp or Lys) was embedded in a TIP3P\cite{mackerell1998all,jorgensen1983comparison} water box with a $40$~\AA~side length. One chloride ion was added to neutralize the box in the lysine case, while one potassium ion and one chloride ion were added to the aspartic acid system. The CHARMM36 force field\cite{charmm36} was used to perform equilibration, and its Lennard-Jones parameters and partial charges were used in the QM/MM simulations.

The systems were first equilibrated at the force field level, then equilibrated at the $\omega$B97X-3c\cite{muller2023omegab97x}/MM level as performed in our previous work\cite{li2024general}. The structures were further equilibrated at the r$^2$SCAN-3c\cite{grimme2021r2scan}/MM level for 5 ps in this work.

We employed the OPES method\cite{invernizzi2020rethinking} to sample proton dissociation from the amino acid side chains, starting from the equilibrated structures. The biased reaction coordinate $\xi$ was the distance between the CEC and either the Lys amine nitrogen or the closest Asp carboxylic oxygen. The CEC is a smooth function of the whole-system's protons and proton-accepting atoms that tracks the position of the excess proton (i.e. \ce{H+}). We adopted the CEC parameters in Ref.~\cite{li2020understanding} for both Lys and Asp in this stage. The distance between the CEC and its closest carboxylic oxygen was approximated by the smooth function
\begin{align}
    \xi = -\frac{1}{40} \ln\big(e^{-20(r_1-r_2)}+e^{-20(r_2-r_1)}\big) + \frac{r_1+r_2}{2}
\end{align}
with $r_1=|\mathbf{r}_\text{CEC}-\mathbf{r}_\text{O1}|$ and $r_2=|\mathbf{r}_\text{CEC}-\mathbf{r}_\text{O2}|$. We used the PLUMED library\cite{tribello_plumed_2014} to implement OPES. The \texttt{BARRIER} parameter was set to 12 kcal/mol for both systems. The Gaussian kernels were deposited every 100 MD steps, and the kernel width was updated adaptively every 200 MD steps. The OPES simulations were run for 50 ps for both systems, and the sampled structures were saved every 50 fs.

In both r$^2$SCAN-3c/MM equilibration and OPES, the QM region was the full amino acid with 27 closest water molecules for Asp, and 28 for Lys. The distance between a water molecule and the Asp is measured by the water oxygen distance from one of the Asp carboxylic oxygen atoms. The water-Lys distance is measured by the water oxygen distance from the amine nitrogen. These distance definitions were also used in the FIRES\cite{rowley2012solvation} restraints (see section~\ref{sec:fires} for details) to keep the QM water close to the solute.

\subsubsection{CM}
\begin{figure}[h]
    \centering
    \includegraphics[width=0.75\linewidth]{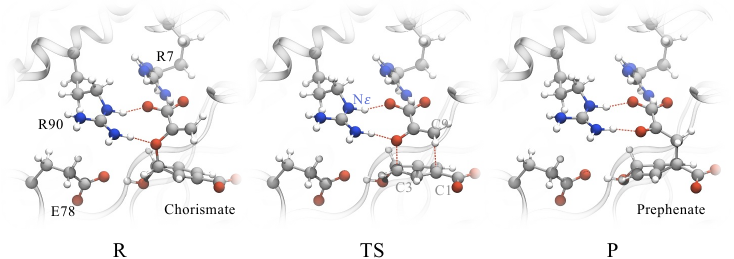}
    \caption{Molecular figures of CM.}
    \label{fig:cm}
\end{figure}

The system and computational setup have been detailed in our previous work\cite{li2025accurate} and are briefly summarized here.

The system comprises a chorismate-bound enzyme solvated in 14821 water molecules with 12 \ce{Na+} ions in a box measuring 79.006~\AA$\times$79.682~\AA$\times$79.030~\AA. We considered both substrate binding modes of CM, the one with one hydrogen bond formed between the substrate and R90, found in PDB 2CHT\cite{chook1993crystal}, and the two-hydrogen-bonded one consistent with PDB 1COM\cite{chook1994monofunctional}. We performed both NEB and SMD to sample representative geometries along the reaction path. We optimized 10 geometries using NEB connecting the reactant to the product state in both binding modes on the $\omega$B97X-3c/MM PES. The SMD biasing the bond-making \ce{C1-C9} distance of the substrate minus the bond-breaking \ce{C3-O} distance (see Fig.~\ref{fig:cm}) was run on a revised $\omega$B97X-3c/MM PES for 7.5 ps, and geometries were saved every 50 fs, resulting in 151 geometries of each binding mode.

\subsection{LNO-CCSD(T) Data Labeling}

\subsubsection{Asp and Lys}

We truncated the r$^2$SCAN-3c/MM OPES sampled structures by keeping the protein side chains with several solvation shells of water. For Asp, we additionally kept the $\alpha$ carbon and capped the dangling bonds with hydrogens placed 1.08~\AA~away from \ce{C_\alpha} along the directions of the original \ce{C-C} bonds. We kept the three solvation shells of the Asp carboxylic proton. The solvation shells were recursively defined as the closest water to the proton. For example, the water whose oxygen was the closest to the carboxylic proton was defined as the first solvation shell, and the two closest water molecules to the first-shell water protons were defined as the second solvation shell. The three solvation shells thus defined included 7 water molecules. For Lys, we truncated at the \ce{C_\alpha-C_\beta} bond and capped \ce{C_\beta} with hydrogen 1.08~\AA~from \ce{C_\beta} along the \ce{C_\beta-C_\alpha} direction. We included two solvation shells of the amine proton with the longest \ce{N-H} bond, and one solvation shell of the two other amine protons, which were 5 water molecules in total.

We selected 100 representative geometries from the full set of 1001 geometries per system. This was done by a k-means clustering analysis in the two-dimensional feature space spanned by their gas-phase energies and $\xi$ values. The gas-phase energies were computed at the MP2/cc-pVTZ level. The energy and $\xi$ values were normalized to have a unit variance before the clustering. The clustering was performed separately for each system, and we selected the 100 geometries with the closest Euclidean distances in the feature space from the clustering centers.

We performed LNO-CCSD(T) calculations on the 100 resulting gas-phase structures of each system to form the initial training set. We direct the readers to Ref.~\cite{zhang2024performant} for a more detailed explanation of our implementation of the LNO-CCSD(T) approach. To formulate the local correlation domains, we first partitioned the system into atomic fragments. The IAOs of each heavy atom with its closest protons were defined as one fragment. The associated MP2 natural orbitals of each fragment's IAOs were truncated according to their occupation numbers. We used a $2\times10^{-5}$ threshold to truncate the virtual LNOs and $2\times10^{-4}$ for the occupied LNOs. The CCSD(T) equations were solved in the space spanned by the LNOs and IAOs of each fragment, and the correlation energies of every fragment were assembled to form a local estimate for the exact CCSD(T) correlation energy. The same space was used to compute an approximate MP2 energy, and its deviation from the exact, global MP2 energy was used as a correction. The final total energy was
\begin{align}
    E^\text{tot}_\text{LNO-CCSD(T)} = E_\text{HF} + E_\text{MP2}^\text{corr,global} - E_\text{MP2}^\text{corr,local} +
    E_\text{CCSD(T)}^\text{corr,local}.
\end{align}
We performed the LNO-CCSD(T) calculations with the cc-pVTZ and cc-pVQZ basis sets, and extrapolated both the HF and correlation energies to the CBS limit using the two-point formula\cite{neese2011revisiting}. The LNO-CCSD(T)/CBS forces were computed through auto differentiation\cite{zhang2024performant}.

\subsubsection{CM}
We performed LNO-CCSD(T) calculations on the 20 NEB geometries of the substrate and the R90 side chain (MM atoms and other parts were removed) to form the training set for DFT. The LNO-CCSD(T) calculation followed the same procedure as for Asp and Lys, except that a tighter threshold was employed ($1\times10^{-5}$ for virtuals and $1\times10^{-4}$ for the occupied). To form the validation set, we performed the same LNO-CCSD(T) calculations on 4 NEB geometries (reactant and transition states, both binding modes) of the substrate, and truncated R90, R7, and E78. The truncation kept E78 as an acetate, the R90 and R7 as methylguanidiniums. All the truncated models were capped by hydrogens following the same protocol as for Asp and Lys.

\section{DFT Parameterization and DFT/MM Data Generation}
\subsection{Asp and Lys}

We re-parameterized the $\omega$B97X-3c functional using the gas-phase LNO-CCSD(T)/CBS data. $\omega$B97X-3c shares the same local density functional component of $\omega$B97X-V but comes with a specifically parameterized double-zeta basis, effective core potential (ECP), and DFT-D4 dispersion correction. In our parameterization, the basis, ECP, range separation parameter $\omega$, and DFT-D4 scaling parameters $s_6$ and $s_9$ were fixed, while all other parameters were optimized. Specifically, we optimized the parameters $c_{\mathrm{x},i}$, $c_{\mathrm{css},i}$, and $c_{\mathrm{cos},i}$ for $i=0$ to 4 (note the original $\omega$B97X-V parameterization sets $c_{\mathrm{x},i}=0$ for $i\geq3$ and $c_{\mathrm{css},i}=c_{\mathrm{cos},i}=0$ for $i\geq2$ but we extended the parameter space), and the D4 parameters $a_1$ and $a_2$. We used the BFGS algorithm to minimize the energy $L^2$ loss, 
\begin{align}
    \frac{1}{\sum_M N_M} |\Delta E_i - \frac{1}{N_M}\sum_j\Delta E_j|^2,
\end{align}
with $\Delta E=E^\text{LNO-CCSD(T)/CBS}-E^\text{DFT}$, and $M$ and $N_M$ take the same meaning as in the MAE.

We first re-parameterized $\omega$B97X-3c on the 100 Asp geometries and tested on the 100 Lys geometry data. In this parameterization, we enforced the short-range exact exchange percentage $c_\mathrm{sr}=1-c_{\mathrm{x},0}$ and the long-range exact exchange $c_\mathrm{lr}=1$. The functional parameterized on Asp energies achieved an energy MAE of 0.47 kcal/mol and a force MAE of 1.1 kcal/mol/\AA~on the Asp training set, with 0.40 kcal/mol (energy) and 0.95 kcal/mol/\AA~(forces) on the independent Lys test set. We then re-trained the DFT on the full Asp and Lys LNO-CCSD(T) data and achieved a 0.43 energy mean absolute training error, and 1.0 kcal/mol/\AA~ force MAE. We denote the reparametrized functional as rev-$\omega$B97X-3c. The fitted parameters are listed in Table~\ref{tab:dftparams}.

We used rev-$\omega$B97X-3c/MM to label the energy and forces of the 1001 Asp/Lys condensed-phase structures sampled from r$^2$SCAN-3c/MM OPES. In these single-point calculations, the QM region was the full amino acid with the same solvation shells of water defined in the gas-phase geometry preparation step. We also performed gas-phase single-point calculations using rev-$\omega$B97X-3c on these structures without the presence of MM charges.

\subsection{CM}

A similar procedure was applied to the 20 CM NEB geometries. We found that a better fit can be achieved by freeing the constraints on $c_\text{sr}$ and $c_\text{lr}$. The resulting parameters are listed in Table~\ref{tab:dftparams}. The fitting error was 0.07 kcal/mol in the energy MAE. We tested the model on larger clusters also containing the truncated R7 and E78 side chains, and the validation error was 0.28 kcal/mol in the energy MAE.

\begin{table}[]
    \centering
    \begin{tabular}{lrr}
    \toprule
    & \multicolumn{1}{c}{Asp/Lys} & \multicolumn{1}{c}{CM} \\
    \midrule
    $\omega$           & 0.30000000  & 0.30000000 \\
    $c_\text{lr}$      & 1.00000000  & 0.38431963 \\
    $c_\text{sr}$      & 0.03999563  & 0.42286641 \\
    $c_{\text{x},0}$   & 0.96000437  & 0.50703519 \\
    $c_{\text{x},1}$   & 3.64182828  & 0.49473559 \\
    $c_{\text{x},2}$   & -0.78053220 & 0.74470496 \\
    $c_{\text{x},3}$   & -1.12103749 & 0.08398222 \\
    $c_{\text{x},4}$   & 0.02291613  & 0.12263558 \\
    $c_{\text{css},0}$ & -1.27911282 & 1.24406763 \\
    $c_{\text{css},1}$ & 0.49534572  & 0.77754377 \\
    $c_{\text{css},2}$ & -0.94957004 & 0.01965672 \\
    $c_{\text{css},3}$ & -1.79543796 & -0.52087769 \\
    $c_{\text{css},4}$ & -1.95602913 & -0.97306147 \\
    $c_{\text{cos},0}$ & -0.50497555 & 1.93939102 \\
    $c_{\text{cos},1}$ & -4.46337449 & -2.13551504 \\
    $c_{\text{cos},2}$ & -1.12640802 & 0.38406090 \\
    $c_{\text{cos},3}$ & 0.31913167  & 0.54762278 \\
    $c_{\text{cos},4}$ & 0.89944993  & 0.46449921 \\
    $a_1$              & 3.54442601  & 0.34530647 \\
    $a_2$              & 4.51860208  & 4.66593285 \\
    \bottomrule
    \end{tabular}
    \caption{Parameters of rev-$\omega$B97X-3c. The definitions of the parameters can be found in Refs.~\cite{mardirossian2014omegab97x,muller2023omegab97x}.}
    \label{tab:dftparams}
\end{table}

The rev-$\omega$B97X-3c/MM parameterized from CM gas-phase data was used in QM/MM to label the energy and forces of the 20 NEB geometries and 302 SMD geometries. The QM region was defined as the substrate and the side chains of R90, R7, and E78, capped with a hydrogen atom at the broken \ce{C_\alpha-C_\beta} bond. We also performed gas-phase rev-$\omega$B97X-3c calculations on these structures.

\section{ML-xTB}
\subsection{Training and Model Details}
\subsubsection{Asp and Lys}
We first fine-tuned the MACE-OFF24(M) model\cite{kovacs2025mace} on the deviation of GFN1-xTB from the rev-$\omega$B97X-3c gas-phase energies and forces. We kept the MACE-OFF24(M) encoder parameters, but initialized the energy decoder (read-out) blocks to predict zero energies and forces. The first 990 geometries of each system were used as the training set, and the last 11 geometries were for validation. The best model was chosen based on the smallest validation error. The training and validation errors take the form of
\begin{align}
    \text{Loss} =\frac{w_E}{r_E^2} \frac{1}{\sum_M N_M}\sum_{M} \sum_{i\in M} |E_i-E_i^\text{ref} - \frac{1}{N_M} \sum_{j\in M} (E_j-E_j^\text{ref})|^2
    + \frac{w_F}{r_F^2} \frac{1}{N_\text{atom}}\sum_a |\mathbf{F}_a-\mathbf{F}^\text{ref}_a|^2
    \label{eq:mlloss}
\end{align}
where $w_E$ and $w_F$ are two unit-less weights, and $r_E$ and $r_F$ are the ranges of energies and forces of the training/validation data sets. Similar to in our definition of MAE, $M$ indexes a system, and $N_M$ is the number of system $M$ data points in the mini-batch, $N_\text{atom}$ is the total number of atoms in the mini-batch, $a$ indexes an atom, and $\mathbf{F}_a$ is the force on that atom.
We used $w_E=0.1$ and $w_F=0.5$ in both the training and validation. We used a batch size of 5 (geometries) in training and 22 for validation. The Adam optimizer was used\cite{kingma2014adam}, with a maximum learning rate $lr_\text{max}$ of $3\times10^{-3}$. The training used a learning rate warmup in the first 3 epochs, followed by a cosine annealing in 80 epochs. The warmup uses a linear schedule starting from $10^{-3}\times lr_\text{max}$.

The resulting model initialized the ML potential part of our ML-xTB/MM architecture (Fig. 1 of the main text). We allowed the model to change all the GFN1-xTB parameters except for those defining the atomic basis: that is $k_\text{s}, k_\text{p}, k_\text{sp}, k_{EN}, a_1, a_2, s_8, K_\text{HH}, k_\text{f},$ and $k_\text{g}$ in  Table 2 of Ref.~\cite{grimme2017robust} and $\Gamma$ in Eq.~2, $\kappa$ in Eq.~5, $k^\text{poly}$ in Eq.~11, $H$ and $k_{\text{CN}}$ in Eq.~12, and $\alpha$ and $Z^\text{eff}$ in Eq.~13. The Table 2 parameters are considered to be ``global" parameters that do not depend on the element type, and we kept them in a look-up table, that is independent of the molecular geometry. We allowed the remaining parameters to be not only element-dependent but also geometry-dependent. We did so by predicting them from the invariant features on each individual atom through atom-wise xTB parameter decoders. The xTB parameter decoders were multi-layer perceptrons with the SiLU activation function and were initialized to predict exactly the same GFN1-xTB parameters. There were no activation functions in the last layer, so the signs of the parameters were not fixed. We also allowed the MM charge magnitudes and radii to be optimized (when computing the QM/MM interaction) instead of taking their values from the classical force field. We restricted the MM parameters to be the same for each element and enforced a water molecule to be charge-neutral. We fixed the \ce{K+} and \ce{Cl-} charges to be +1 and -1 and only optimized their radii. When computing the QM-QM and QM-MM periodic electrostatics in ML-xTB/MM, we used the Coulombic interaction form for Gaussian-distributed charges:
\begin{align}
    E(R_{ij})=\frac{q_i q_j\text{erf}(\eta_{ij} R_{ij})}{R_{ij}}
    \label{eq:coul}
\end{align}
where $\text{erf}$ is the error function and $\eta$ is the averaged hardness of the two charges
\begin{align}
    \eta_{ij} = 2 \Big(\frac{1}{\eta_i}+\frac{1}{\eta_j}\Big)^{-1}
    \label{eq:eta}
\end{align}
and $\eta_i=(1+\kappa_i)\eta_{z_i}$ for xTB shell-resolved charges, while it is a trainable constant for the MM charges. The $\eta_{z_i}$ is the original GFN1-xTB parameter for element type $z_i$ as defined in Eq.~5 of Ref.~\cite{grimme2017robust}~, and $\kappa_i$ is one of the atom-wise decoder model predictions for atom $i$.
We implemented the long-range electrostatic coupling using Ewald summation of the potential (Eq.~\ref{eq:coul}) in dxtb\cite{friede2024dxtb}, rather than the generalized Mataga-Nishimoto-Ohno-Klopman formula ($1/\sqrt{R^2+\eta^2}$) adopted in GFN1-xTB, since the latter is known to yield an ill-defined Ewald potential for an infinite lattice\cite{buccheri2025periodic}.

We trained the full ML-xTB/MM model using the rev-$\omega$B97X-3c/MM data. We used the same loss function (Eq.~\ref{eq:mlloss}), but we defined $N_\text{atom}$ as the number of QM(ML) atoms, although $a$ still ran over all the atoms. The force error would otherwise be trivially small if normalized by the total number of atoms, because of the small QM-MM force errors on distant MM charges. This was due to our model architecture that computes the MM forces using the physical electrostatic coupling instead of being predicted from a neural network.
We trained on the first 990 geometries of each system and validated on the last 11 ones. We used $w_E=0.01$, $w_F=0.99$, and a batch size of 5 for training, and $w_E=0.5$, $w_F=0.5$, and a batch size of 22 for validation. Due to the large system size and relatively small training batch size, the energy fluctuations in a mini-batch were large. As a result, we found that a small $w_E$ was helpful to achieve both smaller energy and force validation errors. We used the Adam optimizer with a maximum learning rate $lr_\text{max}$ of $1\times10^{-4}$. The training used a linear learning rate warmup in the first 3 epochs starting from $10^{-3}\times lr_\text{max}$, followed by a cosine annealing in 80 epochs. The best model was chosen based on the smallest validation error and was used in production runs.

\subsubsection{CM}
The ML-xTB architecture for CM was generally the same as for Asp and Lys, but differed in several aspects. The xTB parameter layer was appended by a Softplus activation to enforce the same signs as the original ones in the GFN1-xTB parameterization. This was found to stabilize the training. The QM-MM electrostatics additionally computed the dipole-charge interactions as
\begin{align}
    E(R_{ij}) = -\frac{\boldsymbol{\mu}_i\cdot \mathbf{R}_{ij} q_j}{R_{ij}^3} \text{erf}(\eta_{ij}R_{ij})
\end{align}
where $\boldsymbol{\mu}$ is the xTB atomic dipoles, and $\eta$ takes the same form as Eq.~\ref{eq:eta}, but with the dipolar hardness for the QM atoms separately parameterized from their charge hardness. The MM charges were not considered trainable but were taken from their classical force field values. The MM radii were kept trainable.

The training and validation procedure was similar to the Asp and Lys case, but used a batch size of 1, a larger maximum learning rate ($5\times10^{-4}$), and elongated the training epoch to 160. 
We included an additional energy offset as a trainable parameter in ML-xTB to facilitate training with energies using a batch size of 1. We excluded the reactant and transition state NEB geometries and 108 transition-state-like SMD geometries from our training set and used them as the validation set.

\subsection{REUS of Proton Dissociation}
We performed ML-xTB/MM REUS to compute the PMF of proton dissociation from the Asp or Lys side chain. The temperature was controlled at 298.15 K via a Langevin thermostat with a 1 fs timestep. The BAOAB time integrator\cite{leimkuhler_robust_2013} was employed for accurate proton motions. We used the same reaction coordinate definition as the QM/MM OPES, but with refined CEC parameters following the constrained DFT approach\cite{li2021using}. The CEC parameters are listed in Table~\ref{tab:cecparams}.

\begin{table}[H]
    \centering
    \begin{tabular}{l|cc}
                       & $k$ (\AA$^{-1}$)  & $\delta_0$ (\AA) \\
    \hline
      \ce{H3O^+-H2O}   & 4.984      & 0                \\
      \ce{Asp-H2O}     & 2.946      & -0.5361          \\
      \ce{Lys-H2O}     & 3.649      & -0.1045          \\
    \end{tabular}
    \caption{The CEC parameters used for production runs. The definition of the parameters is given in Ref.~\cite{li2020understanding}}
    \label{tab:cecparams}
\end{table}

All the umbrella sampling windows were run for 100 ps, and exchanges were attempted at every MD step for all pairs of windows. The bias potential took a harmonic form of $\frac{1}{2}K (\xi-\xi_\text{cen})^2$. The QM(ML) region was defined as the full amino acid plus 64 closest water molecules. The distance between a water molecule and the Asp was defined as the distance between the water oxygen and a virtual atom positioned as $\mathbf{r}_{\text{O}_1}+\mathbf{r}_{\text{O}_2}-\mathbf{r}_\text{C}$ where \ce{O1}, \ce{O2} and \ce{C} are the carboxylic group atoms. The water-Lys distance was defined as the distance between the water oxygen and the side-chain amine nitrogen. A FIRES restraint was applied to keep the QM(ML) definition valid throughout the MD runs. We list the window centers ($\xi_\text{cen}$) and force constants ($K$) of all our simulations in Table~\ref{tab:usparams}.

\begin{table}[H]
    \centering
    \begin{tabular}{cc}
     \hline\hline
     \multicolumn{2}{c}{Asp}                          \\
     \hline     
$\xi_\text{cen}$ (\AA) & $K$ (kcal/mol) \\
 0.250   &    100     \\
 0.500   &    20      \\
 0.750   &    80      \\
 0.875   &    80      \\
 1.000   &    80      \\
 1.125   &    80      \\
 1.250   &    80      \\
 1.500   &    80      \\
 1.750   &    80      \\
 1.875   &    80      \\
 2.000   &    80      \\
 2.250   &    80      \\
 2.500   &    40      \\
 2.750   &    40      \\
 3.000   &    20      \\
 3.500   &    10      \\
 4.000   &    10      \\
 4.500   &    10      \\
 5.000   &    10      \\
     \hline\hline \\
     \hline\hline
     \multicolumn{2}{c}{Lys}                          \\
    \hline    
$\xi_\text{cen}$ (\AA) & $K$ (kcal/mol) \\
  0.250    &  100     \\
  0.500    &  20      \\
  0.750    &  100     \\
  1.000    &  100     \\
  1.250    &  80      \\
  1.500    &  80      \\
  1.750    &  80      \\
  2.000    &  80      \\
  2.250    &  80      \\
  2.500    &  40      \\
  2.750    &  40      \\
  3.000    &  40      \\
  3.250    &  40      \\
  3.500    &  20      \\
  3.750    &  20      \\
  4.000    &  10      \\
  4.500    &  10      \\
     \hline\hline
    \end{tabular}
    \caption{Umbrella sampling settings.}
    \label{tab:usparams}
\end{table}

The proton dissociation PMFs were obtained from WHAM applied to REUS trajectories. The p$K_\text{a}$ of a weak acid is given by
\begin{align}
    \text{p}K_\text{a} = \log{\Big(c_0 \int_0^{\xi^\dagger} \mathrm{d}\xi 4\pi \xi_\infty^2 p(\xi) / p(\xi_\infty)\Big)}
    \label{eq:pka1}
\end{align}
where $c_0=1$ mol/L, $\xi^\dagger$ is the reaction coordinate value that defines a protonated acid, and $\xi^\infty$ is the value when $4\pi\xi^2/p(\xi)$ approaches 1 at the disassociated limit. We used $\xi^\dagger=3$~\AA~for both Asp and Lys and $\xi^\infty=5$~\AA, $4.5$~\AA~for Asp and Lys, respectively. The p$K_\text{a}$ values are not sensitive to these choices within a reasonable range.

\subsection{Conformational Flooding of CM Catalytic Reaction}
We performed ML-xTB/MM conformational flooding to compute the CM catalytic rate constant. The temperature was controlled at 300 K via a Langevin thermostat with a 1 fs timestep. The bias potential (i.e., the flooding potential $V_f$) was set to the negative PMF of the reaction, obtained from previous OPES-flooding simulations\cite{li2025accurate}. The potential was capped at a maximum level $V_f^\text{max}$ to not exceed the reaction free energy barrier, and was restricted to the reactant basin. We used two $V_f^\text{max}$ values, 65 kJ/mol and 59 kJ/mol, and for each of them, we performed 11 flooding runs from previously QM/MM equilibrated structures\cite{li2025accurate}. The flooding simulations were stopped when the system reached the product state, and the stopping time was recorded as $t_f$.

The kinetic rate constant $k_\text{cat}$ from flooding is related to the mean-first-passage-time of the reaction observed in flooding simulations (i.e., $t_f$ in a single run) via
\begin{align}
    k_\text{cat} = \frac{1}{t_f\langle e^{\beta V_f}\rangle_{f,\text{R}}}
\end{align}
We performed the ensemble average over each of the flooding trajectories, and formed estimates of $k_\text{cat}$ when combined with the $t_f$ of each run.

\section{Fine-tuning MACE on the Full System}
We fine-tuned the MACE-OFF23(S) model\cite{kovacs2025mace} on the rev-$\omega$B97X-3c/MM data of Asp and Lys. The training and validation used the same split of ML-xTB/MM, i.e., the first 990 geometries were for training, while the last 11 were for validation. Since the MM water behaves differently from the QM water, we used sulfur and fluorine to differentiate the MM water oxygen and hydrogen from the QM water ones. Due to the low concentration of \ce{K+} and its large distances from Asp/Lys in most geometries, we used bromine, an arbitrary choice, to represent the MM \ce{K+}. The energy read-out layers were initialized to predict zero energies, and we found that it achieved better accuracy than not doing so. The training used the loss Eq.~\ref{eq:mlloss} with $w_E=0$ and $w_F=1$, and a learning rate of $5\times10^{-3}$. The gradients were accumulated in a mini-batch of 16 geometries before updating the model parameters. Other training details were the same as for the training of ML-xTB.

\section{DPRc}
We trained DPRc models on the energy and force differences of GFN1-xTB/MM from rev-$\omega$B97X-3c/MM. The model definition and training parameters for Asp/Lys are detailed in the following deepmd-kit input:
\begin{verbatim}
{
  "model": {
    "type_map": [
      "C",
      "H",
      "O",
      "N",
      "K",
      "Cl",
      "Cm",
      "Hm",
      "Om",
      "Nm",
      "Km",
      "Clm"
    ],
    "type_embedding": {
      "neuron": [
        8
      ],
      "precision": "float32"
    },
    "descriptor": {
      "type": "hybrid",
      "list": [
        {
          "type": "se_a_ebd_v2",
          "sel": [35, 35, 35, 35, 35, 35, 0, 0, 0, 0, 0, 0],
          "rcut_smth": 0.50,
          "rcut": 6.00,
          "neuron": [ 25, 50, 100 ],
          "axis_neuron": 16,
          "precision": "float32",
          "exclude_types": [[6, 6], [6, 7], [6, 8], [6, 9], [6, 10], 
          [6, 11], [7, 7], [7, 8], [7, 9], [7, 10], [7, 11], [8, 8],
          [8, 9], [8, 10], [8, 11], [9, 9], [9, 10], [9, 11], [10, 10],
          [10, 11], [11, 11], [0, 6], [0, 7], [0, 8], [0, 9], [0, 10],
          [0, 11], [1, 6], [1, 7], [1, 8], [1, 9], [1, 10], [1, 11],
          [2, 6], [2, 7], [2, 8], [2, 9], [2, 10], [2, 11], [3, 6],
          [3, 7], [3, 8], [3, 9], [3, 10], [3, 11], [4, 6], [4, 7],
          [4, 8], [4, 9], [4, 10], [4, 11], [5, 6], [5, 7], [5, 8],
          [5, 9], [5, 10], [5, 11]],
          "seed": 1
        },
        {
          "type": "se_a_ebd_v2",
          "sel": [35, 35, 35, 35, 35, 35, 100, 100, 100, 100, 100, 100],
          "rcut_smth": 5.80,
          "rcut": 6.00,
          "neuron": [ 25, 50, 100 ],
          "axis_neuron": 12,
          "set_davg_zero": true,
          "exclude_types": [[0, 0], [0, 1], [0, 2], [0, 3], [0, 4],
          [0, 5],[1, 1], [1, 2], [1, 3], [1, 4], [1, 5], [2, 2],
          [2, 3], [2, 4], [2, 5], [3, 3], [3, 4], [3, 5], [4, 4],
          [4, 5], [5, 5], [6, 6], [6, 7], [6, 8], [6, 9], [6, 10],
          [6, 11], [7, 7], [7, 8], [7, 9], [7, 10], [7, 11],
          [8, 8], [8, 9], [8, 10], [8, 11], [9, 9], [9, 10],
          [9, 11], [10, 10], [10, 11], [11, 11]],
          "precision": "float32",
          "seed": 1
        }
      ]
    },
    "fitting_net": {
      "type": "ener",
      "neuron": [ 240, 240, 240 ],
      "resnet_dt": true,
      "precision": "float32",
      "atom_ener": [ null, null, null, null, null, null,
      0.0, 0.0, 0.0, 0.0, 0.0, 0.0 ],
      "seed": 1
    }
  },
  "learning_rate": {
    "type": "exp",
    "decay_steps": 5000,
    "start_lr": 0.001,
    "stop_lr": 3.51e-8,
  },

  "loss": {
    "type": "ener",
    "start_pref_e": 0.02,
    "limit_pref_e": 1,
    "start_pref_f": 1000,
    "limit_pref_f": 1,
    "start_pref_v": 0,
    "limit_pref_v": 0,
  },    
  
  "training": {
    "numb_steps": 32000,
    "seed": 10,
  },
}
\end{verbatim}
The model definition for CM was the same except for the atom types:
\begin{verbatim}
    "type_map": [
      "C",
      "H",
      "O",
      "N",
      "S",
      "Na",
      "Cm",
      "Hm",
      "Om",
      "Nm",
      "Sm",
      "Nam"
    ],
\end{verbatim}
and training steps:
\begin{verbatim}
  "training": {
    "numb_steps": 50000,
    "seed": 10,
  },
\end{verbatim}
The DPRc training and validation followed the same data split as for ML-xTB/MM.

\section{Computational Details of DFT/MM}
As detailed above, we used DFT-based QM/MM to equilibrate our system, sample training structures, and generate training data. In all the DFT/MM simulations, we used the QM/MM-Multipole approach\cite{li2025accurate} to rigorously treat the long-range QM-QM and QM-MM electrostatics. The method performs exact Ewald summation for all the long-range interactions (charge-charge, charge-dipole, charge-quadrupole, dipole-dipole) while truncating the short-range interactions (charge-octupole and higher order terms) using a real-space cutoff $R_\text{cut}$.

In DFT/MM MD, we determined $R_\text{cut}$ by incrementing its value from 15~\AA~by a step size of 1~\AA~until two consecutive steps both gave a charge-octupole error smaller than $2\times10^{-5}$ Hartree. We used the time-reversible ASPC method\cite{kolafa2004time,kuhne2007efficient} to predict the current MD step 1-RDM for the charge-octupole test and as the initial guess for SCF. For single-point QM/MM energy and force calculations, we used a fixed $R_\text{cut}=25$~\AA.

The force field Lennard-Jones parameters were used for the QM-MM van der Waals interactions. When computing the QM-MM electrostatics, we used Gaussian-distributed MM charges, with magnitudes taken from the force field partial charges, and exponents as the inverse square of the covalent and ionic radii~\cite{pyykko2009molecular,shannon1976revised}.

\section{FIRES Restraint}
\label{sec:fires}
We adopted FIRES in Asp/Lys simulations to restrain the MD sampling in the configurational subspace where the QM water molecules are always the closest to the solute. If the solvent molecules are indistinguishable, the classical partition function of one solute X and $N$ solvent molecules
\begin{align}
    Q = \frac{1}{N!} \int \dd \mathbf{r}_\text{X} \dd \mathbf{r}^N e^{-\beta U(\mathbf{r}_\text{X}, \mathbf{r}_1, \cdots, \mathbf{r}_N)}
\end{align}
can be rewritten as
\begin{align}
    Q = \frac{1}{n! (N-n)!}\int \dd \mathbf{r}_\text{X} \dd \mathbf{r}^N
    e^{-\beta U} \mathbb{I}\big[\max_{i\in\{1,\cdots,n\}}d(\text{X},i)\leq
    \min_{i\in\{n+1,\cdots,N\}}d(\text{X},i)\big]
    \label{eq:firesq}
\end{align}
with $\beta$ being the inverse temperature and $U$ is the potential energy. The indicator function $\mathbb{I}$ in Eq.~\ref{eq:firesq} restricts the integral over the configurational subspace where the first $n$ solvent molecules are the closest $n$ to the solute according to the distance metric $d$. It can be seen that the ensemble average of any time-independent observable is not changed with a modified potential
\begin{align}
    U' = U - \beta^{-1} \ln\mathbb{I}\big[\max_{i\in\{1,\cdots,n\}}d(\text{X},i)\leq
    \min_{i\in\{n+1,\cdots,N\}}d(\text{X},i)\big].
\end{align}
In practice, one approximates the infinite potential $- \beta^{-1} \ln\mathbb{I}\big[\max_{i\in\{1,\cdots,n\}}d(\text{X},i)\leq$\\$
    \min_{i\in\{n+1,\cdots,N\}}d(\text{X},i)\big]$ by a finite potential, e.g.,
\begin{align}
    U^\mathrm{FIRES} = & K^\text{FIRES} \sum_{i=n+1}^N \big( \max(d(\text{X},1),\cdots,d(\text{X},n)) - d(\text{X},i)\big)^2 \times \nonumber\\
    & \mathbb{I}\big[d(\text{X},i)\leq\max(d(\text{X},1),\cdots,d(\text{X},n)\big]
\end{align}

In our QM/MM and QM(ML)/MM simulations, we defined the $n$ water molecules to be the QM or QM(ML) water and used a $K^\text{FIRES}$ of 250 kcal/mol/\AA.

\section{Error Analysis}

\subsection{Convergence of Electronic Structure Theory}
We characterize the convergence of our electronic structure theory in two aspects: (1) error estimates for the LNO truncation in the virtual/occupied spaces, and (2) error estimates for the TZ/QZ to CBS extrapolation.

\subsubsection{Error Estimate of LNO Truncation in LNO-CCSD(T)}
\label{sec:lnoerror}
We performed exact CCSD(T)/cc-pVDZ on the 100 gas-phase geometries of both systems from clustering. The MAE defined by Eq.~\ref{eq:mae} (thus measuring relative energy errors that are relevant to MD) of LNO-CCSD(T) on Asp geometries was 0.07 kcal/mol, and on Lys geometries was 0.02 kcal/mol. We denote this error due to LNO truncation as $\epsilon_\text{LNO}$. Here, we also report the absolute energy MAE, i.e., without the global shifts by mean values as in Eq.~\ref{eq:mae} -- they were 0.56 kcal/mol for Asp and 0.24 kcal/mol for Lys, and thus about ten times larger than the relative energy errors.

We performed the same test for the CM catalytic reaction on 10 NEB geometries (the reactant, product, and transition states, two nearby geometries around the transition state, for each of the two substrate binding modes). The MAE (Eq.~\ref{eq:mae}) of LNO-CCSD(T) was 0.10 kcal/mol compared to the canonical CCSD(T). The absolute energy MAE, without shifting, was 0.47 kcal/mol.

\subsubsection{Error Estimate of Basis Extrapolation}
We compared the TZ/QZ basis extrapolated energies with the QZ energies on the 100 gas-phase geometries. We estimated the basis extrapolation error as half of the energy MAE (Eq.~\ref{eq:mae}) between LNO-CCSD(T)/QZ and LNO-CCSD(T)/CBS. It was 0.08 kcal/mol for Asp and 0.09 kcal/mol for Lys. We denote this error as $\epsilon_\text{CBS}$. The absolute energy difference between QZ and extrapolated CBS was, however, large -- 108 kcal/mol for Asp and 80 kcal/mol for Lys.

We performed the same analysis on the 20 CM geometries (10 NEB geometries for each of the substrate binding modes). The extrapolation error in terms of relative energies (Eq.~\ref{eq:mae}) was small: 0.12 kcal/mol, while the absolute energy difference was 145 kcal/mol.

\subsubsection{Combined Electronic Structure Theory Error}
We estimated our LNO-CCSD(T) energy deviation from the canonical CCSD(T)/CBS as 
\begin{align}
    \epsilon_\text{LNO-CCSD(T)/CBS}=
    \sqrt{
    \epsilon_\text{LNO}^2+
    \epsilon_\text{CBS}^2}.
    \label{eq:ccerror}
\end{align}
The results are 0.11 kcal/mol for Asp, 0.09 kcal/mol for Lys and 0.16 kcal/mol for CM.

\subsection{Statistical Error}
We split the REUS trajectories of Asp and Lys into ten even blocks, applied WHAM to compute ten PMFs from which we computed 10 p$K_\text{a}$'s. We followed the automated equilibration detection approach\cite{chodera2016simple} by throwing out the first $m$ p$K_\text{a}$ samples so that the standard error of mean (SEM) was minimized. We restricted $m$ to be smaller than half of the total sample size for reliable estimates of the SEM. The final p$K_\text{a}$ and its statistical error were reported as the mean and the SEM of the remaining samples.

In CM, we applied exactly the same procedure on 11 $k_\text{cat}$ samples from 11 flooding runs.

\section{Experimental p$K_\text{a}$ Processing}
The p$K_\text{a}$ of Asp and Lys side chains were experimentally measured using potentiometric titrations of model compounds (Ac-Ala-Ala-X-Ala-Ala-NH$_2$ where Ala stands for alanine) in 0.1 M KCl solution at 298.15 K\cite{thurlkill2006pk}. The raw measured values are listed in Table~\ref{tab:exp_pka}. Following the formula 
 in Ref.~\cite{krezel2004formula}, we corrected the raw p$K_\text{a}$ values to be the corresponding D$_2$O values as if they were measured at zero ionic strength using raw glass electrode pH meter readings without isotopic corrections. These corrected results are denoted as p$K_\text{a}^\text{D*}$ in Table~\ref{tab:exp_pka}. We further adjusted the differential response behaviors of glass electrodes to D$^+$ versus H$^+$, following the relation $\text{pD}=\text{pH reading}+0.4$\cite{lumry1951kinetics}. The resulting values (denoted as p$K_\text{a}^\text{D}$ in Table~\ref{tab:exp_pka}) represent the p$K_\text{a}$ of amino acids solvated in heavy water without extra ions, providing more physically relevant quantities for comparison with our theoretical p$K_\text{a}$ calculations based on classical nuclear dynamics at near-zero ionic strength.

 \begin{table}[H]
     \centering
     \begin{tabular}{c|ccc}
          &  raw p$K_\text{a}$ & p$K_\text{a}^\text{D*}$ & p$K_\text{a}^\text{D}$\\
          \hline
    Asp   &  3.67  &  3.44  & 3.84 \\
    Lys   &  10.40 &  10.81 & 11.21
     \end{tabular}
     \caption{Experimental p$K_\text{a}$ values.}
     \label{tab:exp_pka}
 \end{table}


\bibliography{main}